\documentclass[10pt]{article}
\usepackage{natbib}
\usepackage[T1]{fontenc}

\usepackage{amsmath}
\usepackage{amssymb}
\newtheorem{theorem}{Theorem}
\usepackage{graphicx}
\usepackage{epstopdf}

\begin{document}
{\Large
\textbf{Why GPS makes distances bigger than they are}
}
\begin{flushleft}
\bigskip

Peter Ranacher $^{1\ast}$, 
Richard Brunauer$^{2}$, 
Wolfgang Trutschnig $^{3}$
Stefan Christiaan Van der Spek$^{4}$,
and Siegfried Reich $^{2}$

\medskip

{1} Department of Geoinformatics - Z\_GIS, University of Salzburg, Schillerstra{\ss}e 30, 5020 Salzburg, Austria
\\
{2} Salzburg Research Forschungsgesellschaft mbH, Jakob Haringer Stra{\ss}e 5/3, 5020 Salzburg, Austria
\\
{3} Department of Mathematics, University of Salzburg, Hellbrunner Stra{\ss}e 34, 5020 Salzburg, Austria
\\
{4} Delft University of Technology, Faculty of Architecture, Department of Urbanism, Julianalaan 134, 2628BL Delft, The Netherlands

\end{flushleft}

\section*{Abstract}

Global Navigation Satellite Systems (GNSS), such as the Global Positioning System (GPS), are among the most important sensors for movement analysis. GPS is widely used to record the trajectories of vehicles, animals and human beings. However, all GPS movement data are affected by both measurement and interpolation error. In this article we show that measurement error causes a systematic bias in distances recorded with a GPS: the distance between two points recorded with a GPS is -- on average -- bigger than the true distance between these points. This systematic `overestimation of distance' becomes relevant if the influence of interpolation error can be neglected, which is the case for movement sampled at high frequencies. We provide a mathematical explanation of this phenomenon and we illustrate that it functionally depends on  the autocorrelation of GPS measurement error ($C$). We argue that $C$ can be interpreted as a quality measure for movement data recorded with a GPS. If there is strong autocorrelation any two consecutive position estimates have very similar error. This error cancels out when average speed, distance or direction are calculated along the trajectory.  

Based on our theoretical findings we introduce a novel approach to determine $C$ in real-world GPS movement data sampled at high frequencies. We apply our approach to a set of pedestrian and a set of car trajectories. We find that the measurement error in the data is strongly spatially and temporally autocorrelated and give a quality estimate of the data. Finally, we want to emphasize that all our findings are not limited to GPS alone. The systematic bias and all its implications are bound to occur in any movement data collected with absolute positioning if interpolation error can be neglected.

\section{Introduction}

Global Navigation Satellite Systems (GNSSs), such as the Global Positioning System (GPS), have become essential sensors for collecting the movement of objects in geographical space. In movement ecology, GPS tracking is used to unveil the migratory paths of birds \citep{higuchi_pierre_2005}, elephants \citep{douglas-hamilton_etal_2005} or roe deer \citep{andrienko_etal_2011a}. In urban studies, GPS movement data help detecting traffic flows \citep{zheng_etal_2011a} and human activity patterns in cities \citep{vanderspek_etal_2009}. In transportation research, GPS allows monitoring of intelligent vehicles \citep{zito_etal_1995} and mapping of transportation networks \citep{mintsis_etal_2004}, to name but a few application examples.

Movement recorded with a GPS is commonly stored in form of a trajectory. A  trajectory $\tau$ is an ordered sequence of spatio-temporal positions: $\tau ={< (\boldsymbol{P_1},t_1),...,(\boldsymbol {P_n},t_n)>}$, with $t_1 < ... < t_n$ \citep{gueting_schneider_2005}. The tuple $(\boldsymbol{P},t)$ indicates that the moving object was at a position $\boldsymbol{P}$ at time $t$. In order to represent the continuity of  movement, consecutive positions $(\boldsymbol{P_i},t_i)$ and $(\boldsymbol{P_{j}},t_{j})$ along the trajectory are connected by an interpolation function \citep{macedo_etal_2008}. 

However, although satellite navigation provides global positioning at an unprecedented accuracy, GPS trajectories remain affected by errors. There are two types of error inherent in any kind of movement data, measurement error and interpolation error \citep{schneider_1999}, and these errors inevitably also affect trajectories recorded with a GPS.

\begin{itemize}

\item Measurement error refers to the impossibility of determining the actual position $(\boldsymbol{P},t)$ of an object due to the limitations of the measurement system. In the case of satellite navigation, it reflects the spatial uncertainty associated with each position estimate. 

\item Interpolation error refers to the limitations on interpolation representing the actual motion between consecutive positions $(\boldsymbol{P_i},t_i)$  and $(\boldsymbol{P_{j}},t_{j})$. This error is influenced by the temporal sampling rate at which a GPS records the positions. 

\end{itemize}

Measurement and interpolation errors cause the movement recorded with a GPS to differ from the actual movement of the object. This needs to be taken into account in order to achieve meaningful results from GPS data. 

In this article we focus on GPS measurement error in movement data. We show that measurement error causes a systematic overestimation of distance. Distances recorded with a GPS are -- on average -- always bigger than the true distances travelled by a moving object, if the influence of interpolation error can be neglected. In practice, this is the case for movement recorded at high frequencies. We provide a rigorous mathematical explanation of this phenomenon. Moreover, we show that the overestimation of distance is functionally related to the spatio-temporal autocorrelation of GPS measurement error. We build on this relationship and provide a novel methodology to assess the quality of GPS movement data. Finally we demonstrate our method on two types of movement data, namely the trajectories of pedestrians and cars.

Section~\ref{sec:related work} introduces relevant work from previously published literature. Section \ref{sec:overestimation}  provides a mathematical explanation of why GPS measurement error causes a systematic overestimation of distance.  Section~\ref{sec:autocorr} shows how this overestimation can be used to reason about the spatio-temporal auto-correlation of measurement error. Section \ref{sec:experiment} describes the experiment and presents our experimental results, section \ref{sec:discussion} discusses the results.\\ 

\section{Related work} \label{sec:related work}

Since GPS data have become a common component of scientific analyses its quality parameters have received considerable attention. These parameters include the accuracy of the estimated position, the availability and the update rate of the GPS signal, as well as the continuity, integrity, reliability and coverage of the service \citep{hofmann-wellenhof_etal_2003}. The accuracy of the estimated position (i.e. the expected conformance of a position provided with a GPS to the true position, or the anticipated measurement error) is clearly of utmost importance. Measurement error and its causes, influencing factors, and scale have been extensively discussed in published literature: measurement error has been shown to vary over time \citep{olynik_2002} and to be location-dependent. Shadowing effects, for example due  to canopy cover, have a significant influence on its magnitude \citep{deon_etal_2002}. Measurement error is both random and caused by external influences, as well as systematic and caused by the system's limitations \citep{parent_etal_2013}. 

Measurement error is the result of several influencing factors. According to \cite{langley_1997}, these include:

\begin{itemize}
\parskip 10pt

\item Propagation delay:  atmospheric variations can affect the speed of the GPS signal and hence the time that it takes to reach the receiver;

\item drift in the GPS clock: a drift in the on-board clocks of the different GPS satellites causes them to run asynchronously with respect to each other and to a reference clock;

\item Ephemeris error: imprecise satellite data and incorrect  physical models affect estimations of the true orbital position of each GPS satellite \citep{colombo_1986}; 

\item Hardware error: the GPS receiver, being as fault-prone as any other measurement instrument, produces an error when processing the GPS signal;

\item Multipath propagation: infrastructure close to the receiver can reflect the GPS signal and thus prolong its travel time from the satellite to the receiver;

\item Satellite geometry: an unfavourable geometric constellation  of the satellites reduces the accuracy of positioning results.
\end{itemize}

There are several quality measures to describe GPS measurement error, the most common being the \emph {95 $\%$ radius} ($R95$), which is defined as the radius of the smallest circle that encompasses 95~$\%$ of all position estimates \citep{chin_1987}.  The official GPS Performance Analysis Report for the Federal Aviation Administration issued by the \cite{gps_report_2013} states that the current set-up of the GPS allows to measure a spatial position with an average $R95$ of slightly over 3 meters using the Standard Positioning Service (SPS). 
The values  in the report were, however, obtained from reference stations that were equipped with high quality receivers and had unobstructed views of the sky. It is reasonable to assume that the accuracy would be reduced in other recording environments, as measurement error depends to a considerable extent on the receiver, as well as on the geographic location \citep{gps_report_2013, langley_1997}. This assumption is supported by published literature on GPS accuracy in forests \citep{sigrist_etal_1999} and on urban road networks  \citep{modschnig_etal_2006}, as well as on the accuracies of different GPS receivers \citep{wing_etal_2005, zandbergen_2009}. On the other hand, the accuracy of GPS can be increased using  differential global positioning systems (DGPS) such as the European Geostationary Navigation Overlay Service (EGNOS). DGPS  estimate and correct the propagation delay in the ionosphere, thus yielding higher position accuracies \citep{hofmann-wellenhof_etal_2003}. 

A detailed overview of current GPS accuracy is provided in the quarterly GPS Performance Analysis Report for the Federal Aviation Administration. A good introduction to the GPS in general, and to its error sources and quality parameters in particular, has been provided by \cite{hofmann-wellenhof_etal_2003}.\\

The above-mentioned research has mainly focused on describing and understanding GPS measurement error. In addition to this,  filtering and smoothing approaches have been proposed for recording movement data, in order to reduce the influence of errors on movement trajectories. A concise summary of these approaches can be found in  \cite{parent_etal_2013}. \cite{jun_etal_2006} tested smoothing methods that best preserve travelled distance, speed, and acceleration. The authors found that Kalman filtering resulted in the least difference between the true movement and its representation.

\section{GPS measurement error causes a systematic overestimation of distance} \label{sec:overestimation}

A GPS measurement consists of a spatial component (i.e. latitude~$\phi$, longitude~$\lambda$) and a temporal component (i.e. a time stamp $t$).  In this article we will mainly focus on the spatial component. 

The GPS uses the \emph{World Geodetic System 1984 (WGS84)} as a coordinate reference system. For reasons of simplicity it is preferable to transform the GPS measurements to a Cartesian map projection, such as the Universal Transversal Mercator (UTM). A transformation from an ellipsoid (WGS84) to a Cartesian plane (UTM) leads to a distortion of the original trajectories \citep{hofmann-wellenhof_etal_2003}.  For vehicle, pedestrian, or animal movements  consecutive positions along a trajectory are usually sampled in intervals ranging from seconds to minutes. Thus, these positions are very close together in space so that the distortion is insignificant for most practical applications. Hence, for all the following consideration we assume that the movement is recorded in UTM. 

Very generally,  a spatial position in UTM is a two-dimensional coordinate

\begin{equation}
{\boldsymbol{P}} = \left(
\begin{array}{c}
x\\
y\\
\end{array}
\right) ,
\end{equation} 

where $x$ is the metric distance of the position from a reference point in eastern direction and $y$ in northern direction. If a moving object is recorded at position ${\boldsymbol{P}}$ with a GPS, the position estimate
$ \boldsymbol{P^m} =  (x^m,   y^m)$  is affected by measurement error. The relationship between the true position and its estimate is very trivially

\begin{equation}
\boldsymbol {P^m} = \boldsymbol{P}+\boldsymbol{{\varepsilon_P}} \text{,}
\end{equation}
where $\boldsymbol{{\varepsilon_P}}$ is the horizontal measurement error expressed as a vector in the horizontal plane. $\boldsymbol{{\varepsilon_P}}$ is drawn from $\boldsymbol{{\mathcal{E}_P}}$, the distribution of measurement error at $\boldsymbol{{P}}$. We adopt the convention used by \cite{codling_etal_2008} to denote random variables with upper case letters and their numerical values with lower case letters.

\begin{figure}[]
	\centering
    \includegraphics[scale=0.4]{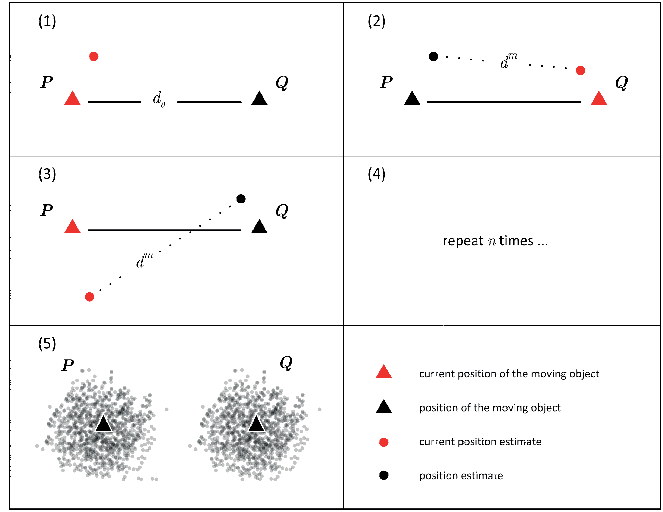} 
	\caption{A moving object equipped with a GPS travels between two arbitrary positions}  
	\label{fig:thought_experiment}
\end{figure}

We now provide a detailed mathematical explanation of why measurement error causes a systematic overestimation of distance in trajectories, if interpolation error can be neglected. 
Figure~\ref{fig:thought_experiment} illustrates the problem statement in a simplified form. Consider a moving object equipped with a GPS. The moving object travels between two arbitrary positions $\boldsymbol{P}$ and $\boldsymbol{Q}$. Let $d_0 = d(\boldsymbol{P},\boldsymbol{Q})$ denote the Euclidean distance between these, henceforth referred to as reference distance. The object always moves along a straight line, consequently interpolation error can be neglected. The movement of the object can be described by the following five steps, which correspond to the subplots in Figure~\ref{fig:thought_experiment}.

\begin{enumerate}
\item {

The moving object starts at $\boldsymbol{P}$. The GPS obtains the position estimate $\boldsymbol{P^m}$ with measurement error $\boldsymbol{{\varepsilon_P}}$, which is drawn from $\boldsymbol{{\mathcal{E}_P}}$.}
\item { 
The moving object travels to $\boldsymbol{Q}$. The GPS obtains the position estimate $\boldsymbol{Q^m}$ with measurement error $\boldsymbol{{\varepsilon_Q}}$, which is drawn from $\boldsymbol{{\mathcal{E}_Q}}$. The distance between the two position estimates is calculated:  ${d^m} = d(\boldsymbol{P^m},\boldsymbol{Q^m})$.} 
\item{ 
The moving object returns to $\boldsymbol{P}$. The GPS obtains a position estimate and a new ${d^m}$ is calculated.}
\item {Steps 2 and 3 are repeated $n$ times, where $n$ is an infinitely large number.}
\item {After $n$ repetitions, the position estimates scatter around  $\boldsymbol{P}$ and $\boldsymbol{Q}$ with measurement error $\boldsymbol{{\mathcal{E}_P}}$ and $\boldsymbol{{\mathcal{E}_Q}}$.}

\end{enumerate}

We claim 
that measurement error propagates to the expected measured distance 
$\mathbb{E}(d^m)$ and to the expected squared measured distance 
$\mathbb{E} (d_2^m)$ between the position estimates. More specifically, measurement error yields $\mathbb{E} (d^m) >d_0$ as well as $\mathbb{E}(d_2^m) >d_0^2$. 

We are now going to 
rigorously prove this claim. 
To do so, we simplify notation, write $\boldsymbol{{\mathcal{E}_P}}=(X_1,Y_1)$ as well as
$\boldsymbol{{\mathcal{E}_Q}}=(X_2,Y_2)$, and assume that there is no systematic bias, i.e. we have 
$\mathbb{E}(X_1)=\mathbb{E}(X_2)=\mathbb{E}(Y_1)=\mathbb{E}(Y_2)=0$. Since neither translations nor rotations affect
distances between points we may, without loss of generality, consider $\boldsymbol{P}=(0,0)$ and $\boldsymbol{Q}=(d_0,0)$.
Since linear transformations (like rotations) preserve expectation, rotating errors with expectation zero results in 
errors having expectation zero too. Having this we can now formulate the following first result for the 
expected squared distance $\mathbb{E}(d^2(\boldsymbol{P^{m}}, \boldsymbol{Q^{m}}))$.
For mathematical background we refer to \cite{klenke_2013}. Notice that no assumptions 
(like absolute continuity or normality) about the underlying error distributions are needed, i.e. the result holds in full generality.

\begin{theorem}  \label{th:overestimation}
Suppose that $d_0>0$, that $\boldsymbol{P}=(0,0)$, and that $\boldsymbol{Q}=(d_0,0)$. Let 
$X_1,X_2$ both have distribution function $F$ and variance $\sigma_X^2$, and $Y_1,Y_2$ both have distribution function $G$ 
and variance $\sigma_Y^2$. Furthermore assume that
$\mathbb{E}(X_1)=\mathbb{E}(X_2)=\mathbb{E}(Y_1)=\mathbb{E}(Y_2)=0$. Then the following two conditions are equivalent:
\begin{enumerate}
\item $\mathbb{E}(d_2^m)= \mathbb{E}(d^2(\boldsymbol{P^{m}}, \boldsymbol{Q^{m}})) > d_0^2$
\item $\min\{Cov(X_1,X_2), Cov(Y_1,Y_2)\} <1 $
\end{enumerate}  
In other words: the expected squared distance $\mathbb{E}(d_2^m)$ is strictly greater than $d_0^2$ unless the errors fulfil $X_1=X_2$ 
and $Y_1=Y_2$ with probability one (which describes the situation of always having identical errors in 
$\boldsymbol{P}$ and $\boldsymbol{Q}$).  
\end{theorem} 
\textbf{Proof:} Calculating $\mathbb{E}(d^2(\boldsymbol{P^{m}}, \boldsymbol{Q^{m}}))$ and using the fact that 
$\textit{Cov}(X_1,X_2)\leq \sigma_X^2$ and $\textit{Cov}(Y_1,Y_2)\leq \sigma_Y^2$ directly yields
\begin{eqnarray} \label{eq:overestimation}
\mathbb{E}(d^2(\boldsymbol{P^{m}},\boldsymbol{Q^{m}})) &=& \mathbb{E}(d_0+X_2-X_1)^2 + \mathbb{E}(Y_2-Y_1)^2 \nonumber \\
 &=& d_0^2 + \mathbb{E}(X_2-X_1)^2 + \mathbb{E}(Y_2-Y_1)^2 =  d_0^2 + \textit{Var}(X_2-X_1) + \textit{Var}(Y_2-Y_1) \nonumber \\
 &=& d_0^2 + 2 \sigma_X^2 + 2 \sigma_Y^2 - 2 \textit{Cov}(X_1,X_2) - 2 \textit{Cov}(Y_1,Y_2) \\
 &\geq& d_0^2.\nonumber
\end{eqnarray}
Having this it follows immediately that $\mathbb{E}(d^2(\boldsymbol{P^{m}},\boldsymbol{Q^{m}}))=d_0^2$ if and only if
$\textit{Cov}(X_1,X_2)= \sigma_X^2$ and $\textit{Cov}(Y_1,Y_2)=\sigma_Y^2$, which in turn is equivalent to the fact that 
$X_1=X_2$ and $Y_1=Y_2$ holds with probability one. $\blacksquare$   

In general one is, however, interested in the expected distance $\mathbb{E} (d^m):= \mathbb{E}(d(\boldsymbol{P^{m}}, \boldsymbol{Q^{m}}))$ 
and not in the expected squared distance. Since, in general, $\mathbb{E}(Z^2)>d_0^2$ need not imply 
$\mathbb{E}(\vert Z \vert)>d_0$ for arbitrary random variables $Z$, a different method is used to prove the 
following main result: 
\begin{theorem}  \label{th:overestimation2}
Suppose that the assumptions of Theorem \ref{th:overestimation} hold. Then the following two conditions are equivalent:
\begin{enumerate}
\item $\mathbb{E}(d^m)= \mathbb{E}(d(\boldsymbol{P^{m}}, \boldsymbol{Q^{m}})) > d_0$
\item $\max\big\{\mathbb{P}(Y_1\not =Y_2), \mathbb{P}(X_2-X_1<-d_0)\big\} >0$
\end{enumerate}  
In other words: the expected distance $\mathbb{E}(d^m)$ is strictly greater than the true distance $d_0$ unless the errors fulfil $Y_1=Y_2$
with probability one and $\mathbb{P}(X_2-X_1<-d_0)=0$ holds. 
\end{theorem} 
\textbf{Proof:} Obviously we have 
\begin{equation}\label{temp}
\sqrt{(d_0+X_2-X_1)^2 + (Y_2-Y_1)^2} \geq \vert d_0+X_2-X_1  \vert 
\end{equation}
Setting $Z:=X_2-X_1$ implies $\mathbb{E}(Z)=0$. Assume now that $\mathbb{P}(Z<-d_0)>0$ holds. Then the desired inequality follows 
immediately from 
\begin{eqnarray}\label{temp2}
\mathbb{E}\vert Z + d_0 \vert &=& \int_{\mathbb{R}} \vert z+ d_0 \vert \, d \mathbb{P}^Z = 
\int_{[-d_0,\infty]} (z+d_0) \,d \mathbb{P}^Z + \int_{(-\infty,-d_0)} -(z+d_0) \,d \mathbb{P}^Z \nonumber \\
&=& \underbrace{\int_{\mathbb{R}} (z+d_0) \,d \mathbb{P}^Z}_{=d_0} \,+ \,\,
  \underbrace{(-2) \int_{(-\infty,-d_0)} (z+d_0) \,d \mathbb{P}^Z}_{>0}\\
&>& d_0. \nonumber
\end{eqnarray}   
In case we have $\mathbb{P}(Z<-d_0)=0$ but $\mathbb{P}(Y_1\not=Y_2)>0$ holds, then 
Inequality~\ref{temp} is strict with positive probability so we get 
$$
\mathbb{E}(d^m)=\mathbb{E}\Big(\sqrt{(d_0+X_2-X_1)^2 + (Y_2-Y_1)^2} \Big)>\mathbb{E}(\vert d_0 +X_2-X_1  \vert)= \mathbb{E}(\vert Z + d_0 \vert)  = d_0. 
$$
Altogether this shows that the second condition of Theorem \ref{th:overestimation2} implies the first one. \\
To prove the reverse implication, assume that $\max\big\{\mathbb{P}(Y_1\not =Y_2), \mathbb{P}(X_2-X_1<-d_0)\big\} =0$.
Then, firstly, the left and the right hand-side of Inequality~\ref{temp} coincide with probability one, so 
$\mathbb{E}(d^m)=\mathbb{E}\big(\sqrt{(d_0+X_2-X_1)^2 + (Y_2-Y_1)^2} \big)=\mathbb{E}(\vert d_0 +X_2-X_1  \vert)$ holds. 
And secondly, directly applying Equality~\ref{temp2} yields $\mathbb{E}(\vert Z + d_0 \vert)  = d_0$, which finally shows
$\mathbb{E}(d^m)= d_0$. $\blacksquare$ 
\newtheorem{rem} {Remark}
\begin{rem}  \label{corr:overestimation}
\emph{
It is worth mentioning that Theorem \ref{th:overestimation2} has several interesting (and partially surprising)
consequences: Whenever the errors in $x$-direction are unbounded (like in the case of normal distributions) the expected distance
is always strictly greater than the true distance $d_0$. The same holds whenever the errors $Y_1$ and $Y_2$ in $y$-direction
do not always coincide -- a very realistic assumption for GPS trajectories.  
}
\end{rem}

We want to underline that Theorem~{\ref{th:overestimation} and \ref{th:overestimation2} hold in full generality for arbitrary distributions of GPS measurement error. Although GPS measurement error is often assumed to have a bivariate normal distribution and to be independent in both the $x$- and $y$-direction \citep{jerde_visscher_2005, bos_etal_2007}, \cite{chin_1987} puts forward convincing arguments why this is very likely not the case. Hence, the general validity of our findings is relevant. 

For reasons of simplicity, we have assumed that $\boldsymbol{{\mathcal{E}_P}}$ and $\boldsymbol{{\mathcal{E}_Q}}$  follow the same distribution function and that there is no systematic bias, i.e. $\boldsymbol{{\mathcal{E}_P}}$ is centred around $\boldsymbol{P}$ and $\boldsymbol{{\mathcal{E}_Q}}$ around $\boldsymbol{Q}$. This assumption is generally acknowledged for in the literature. It builds, for example,  the basis for algorithms to extract road maps from GPS tracking data  \citep[e.g.][]{wang_etal_2014}. Roads are assumed to be located where the density of the GPS measurements is the highest. Also Figure~\ref{fig:collage} shows that this assumption is indeed realistic for real-world GPS data.  However, even a systematic bias does not necessary restrict the validity of our argument. Let us assume that $\mathbb{E}(X_1)=\mathbb{E}(X_2) \neq 0$ and $\mathbb{E}(Y_1)=\mathbb{E}(Y_2)\neq 0$, i.e. the mean of the error distribution has shifted away from $\boldsymbol{P}$ and $\boldsymbol{Q}$ respectively. As the shift is the same for  $\boldsymbol{{\mathcal{E}_P}}$  and $\boldsymbol{{\mathcal{E}_Q}}$, the influence on distance calculations cancels out, Theorem~{\ref{th:overestimation} and \ref{th:overestimation2} still hold. The validity of our proof is restricted only if $\mathbb{E}(X_1) \neq \mathbb{E}(X_2)$ or $\mathbb{E}(Y_1) \neq \mathbb{E}(Y_2)$. This implies that the mean of the error distribution changes abruptly between $\boldsymbol{P}$ and $\boldsymbol{Q}$. As -- in practice -- $\boldsymbol{P}$ and $\boldsymbol{Q}$ are very close in space, this scenario is not realistic for GPS measurement error.

\section{How big is the overestimation of distance and why is this relevant?} \label{sec:autocorr}

In the previous Section we proved that distances recorded with a GPS are on average bigger than the distances travelled by a moving object, if interpolation error can be neglected. In this Section we provide an equation for $\textit{OD}$, the expected overestimation of distance. Moreover, we identify  three parameters that influence the magnitude of $\textit{OD}$. First, let us define $\textit{OD}$ with the help of Equation~\ref{eq:overestimation}: 
\[
\textit{OD} = \mathbb{E}(d^m_2)^{\frac{1}{2}} - d_0  = (d^2_0 + 2 \sigma_X^2 + 2 \sigma_Y^2 - 2 \textit{Cov} (X_1,X_2) - 2 \textit{Cov} (Y_1,Y_2)) ^{\frac{1}{2}} - d_0.
\]
From this follows that $\textit{OD}$ is a function of three parameters: 

\begin{enumerate}

\item{$d_0$, the reference distance between $\boldsymbol{P}$ and $\boldsymbol{Q}$}\\

\item{$\textit{Var}_{\textit{gps}} = 2 \sigma_X^2 + 2 \sigma_Y^2$, a term for the variance of GPS measurement error}\\

\item{$C = 2 \textit{Cov} (X_1,X_2) - 2 \textit{Cov} (Y_1,Y_2)$, a term for the auto-correlation of GPS measurement error. $C$ expresses the similarity of any two consecutive position estimates. If $C$ is big, consecutive position estimates are affected by similar GPS measurement error (see also Figure~\ref{fig:influencing_parameters}}).
\end{enumerate}

We can now simplify notation and write 

\begin{equation} 
\label{eq:simplified_OD}
\textit{OD} =  (d^2_0 + \textit{Var}_{\textit{gps}} - C) ^{\frac{1}{2}} - d_0.
\end{equation}

The influence of the three parameters on $\textit{OD}$ is further illustrated in Figure~\ref{fig:influencing_parameters}. $\textit{OD}$ is small if the reference distance is big, the variance of GPS measurement error is small and the error has high positive autocorrelation. $\textit{OD}$ is big if the reference distance is small, the variance of  GPS measurement error is big and the error has high negative autocorrelation.

\begin{figure}[]
	\centering
   \includegraphics[scale=0.5]{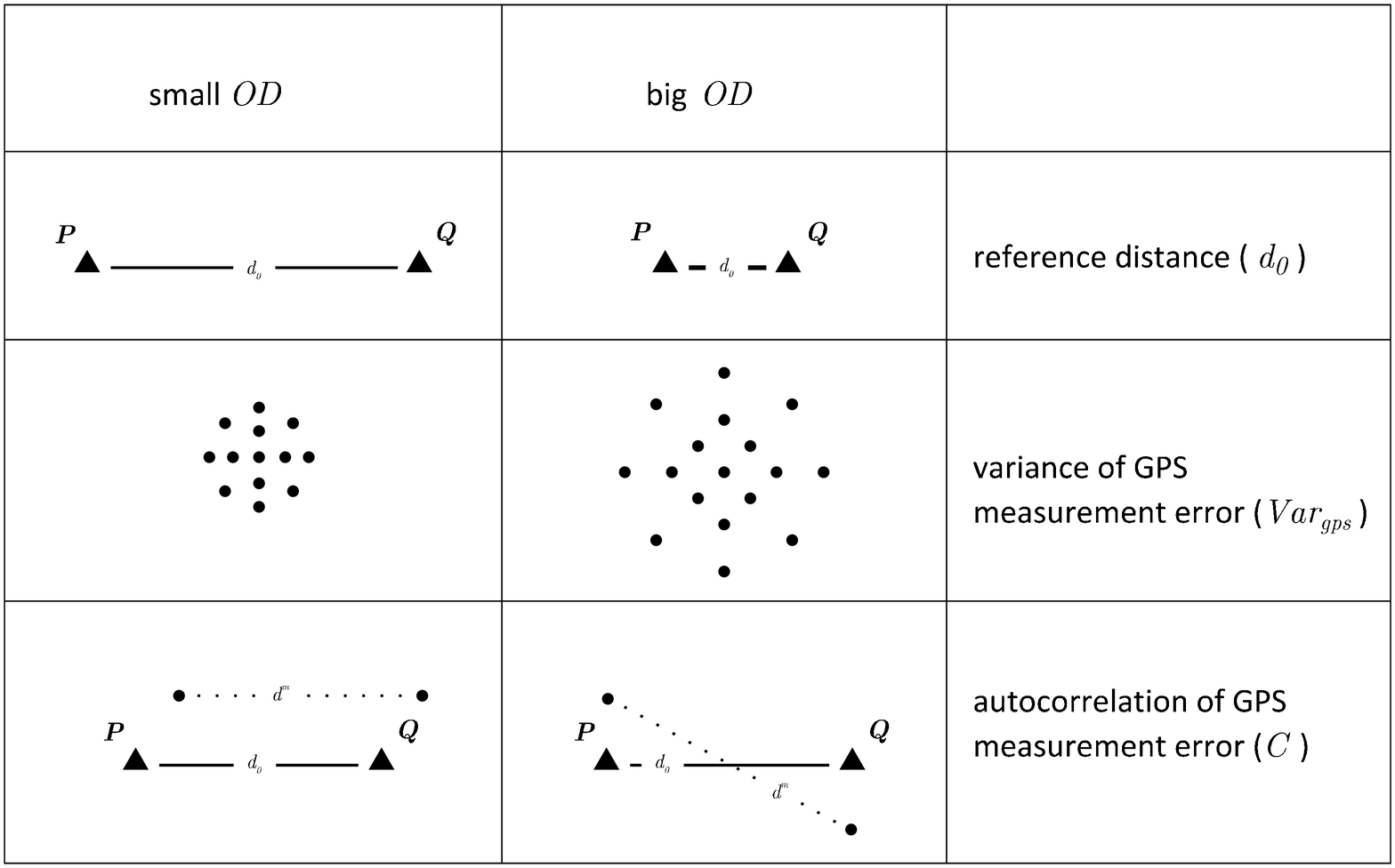} 
	\caption{Overestimation of distance ($\textit{OD}$) and its influencing parameters.}  
	\label{fig:influencing_parameters}
\end{figure}

To understand the magnitude of $\textit{OD}$ in real-world GPS data, let us assume for a moment that  there is no autocorrelation of GPS measurement error, i.e. $C =0$. Moreover, let us assume that the variance of error is the same in $x$- and $y$- directions, i.e. $\sigma^2 = \sigma_X^2 = \sigma_Y^2$ and $\textit{Var}_{\textit{gps}} = 4 \sigma^2$. We can now visualize the relationship between $\textit{OD}$, $d_0$ and $\sigma$. Figure~\ref{fig:OD_compared}~a} shows that $\mathit{OD}$ increases as the spread of GPS measurement error ($\sigma$) increases; $d_0$ is assumed to be constant. For a constant $d_0$ of $5{m}$, for example, and $\sigma =2 \rm {m}$  the overestimation of distance roughly equals $2\rm{m}$ (yellow line). When $\sigma$ increases to $4 \rm {m}$ the overestimation of distance increases to $4 \rm {m}$. Figure~\ref{fig:OD_compared}~b shows that $\mathit{OD}$ decreases as $d_0$ increases; $\sigma$ is assumed to be constant.  For a constant $\sigma$ of $3 \rm {m}$, for example, and $d_0 = 5{m}$ the overestimation of distance equals around $3\rm{m}$ (black line). When $d_0$ increases to $10{m}$ the overestimation of distance decreases to $2 \rm {m}$.

\begin{figure}[h]
	\centering
    \includegraphics[scale=0.4]{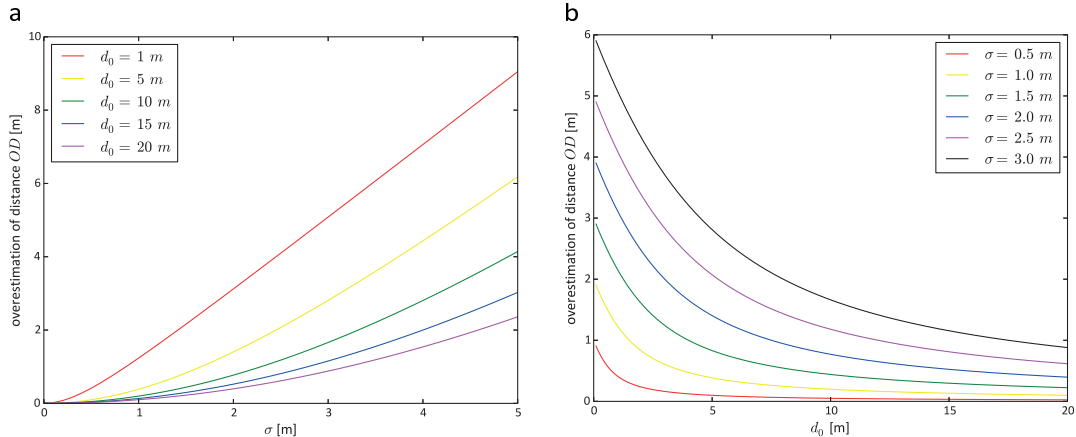} 
	\caption{The overestimation of distance $(\mathit{OD}$) increases as the spread of GPS measurement error ($\sigma$) increases, the reference distance ($d_0$) is constant (a); $\mathit{OD}$ decreases as $d_0$ increases $\sigma$ is constant (b).}
	\label{fig:OD_compared}
\end{figure}

Remember that Figure~\ref{fig:OD_compared}} shows the influence of $\textit{Var}_{\textit{gps}}$ if there is no autocorrelation of GPS measurement error. This is not very realistic for real world GPS data. In fact, \cite{el-rabbany_kleusberg_2003, wang_etal_2002} and \cite{howind_etal_1999} show that GPS is affected by both spatial and temporal autocorrelation. This means that position estimates taken close in space and in time tend to have similar error. 

How big is the autocorrelation of GPS measurement error? Let us reformulate Equation~\ref{eq:simplified_OD} and solve for $C$:  

\begin{equation}
\label{eq:C}
C = d_0^2 - \overbrace{•(\textit{OD} + d_0)^2} ^{\mathbb{E}(d^m_2)} + \textit{Var}_{\textit{gps}}.
\end{equation}

This implies that we can calculate the autocorrelation of GPS measurement error if $\mathit{OD}$, $\textit{Var}_{\textit{gps}}$ and $d_0$ are known. Things become interesting if we consider what autocorrelation really means in the context of GPS positioning. In Figure~\ref{fig:influencing_parameters}, in the bottom left cell, the position estimates $\boldsymbol{P^m}$ and $\boldsymbol{Q^m}$ are highly auto-correlated and, hence, very similar. This leads to the effect that $d^m$ is very similar to $d_0$. In fact, this applies not only to distance, but to other movement parameters as well.  Direction, speed, acceleration or turning angle must all be similar to the `true' movement of the object if they are derived from highly auto-correlated GPS position estimates. Consequently, $C$ describes how well a GPS captures the movement of an object, if interpolation error can be neglected.  Or in other words, $C$ is a quality measure for GPS movement data.

\section{Assessing the quality of GPS movement data} \label{sec:experiment}

\begin{figure}[]
	\centering
   \includegraphics[scale=0.33]{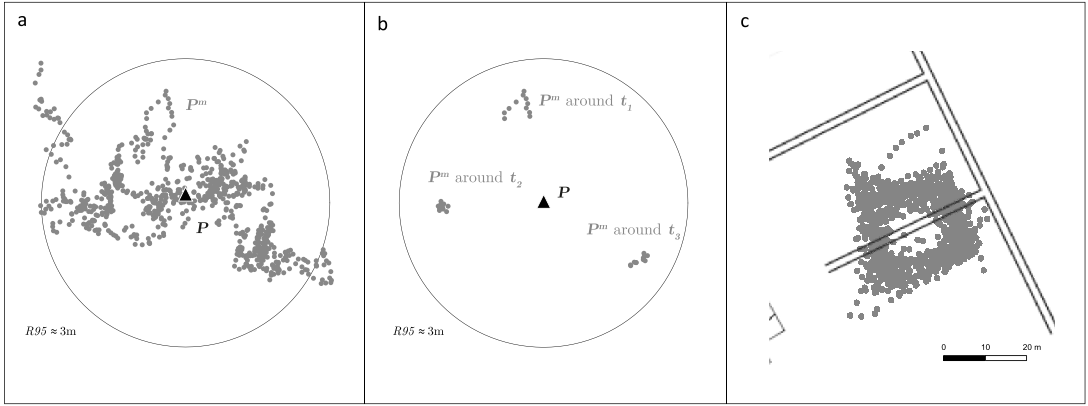} 
	\caption{The distribution of GPS measurement error at position $\boldsymbol{P}$ (a). Revealing the temporal autocorrelation of GPS measurement error (b). The movement of a pedestrian around a reference course (c).}  
	\label{fig:collage}
\end{figure}

Real world GPS data are affected by spatial as well as temporal autocorrelation \citep{el-rabbany_kleusberg_2003, wang_etal_2002, howind_etal_1999}.  Spatial autocorrelation implies that GPS measurement error is not independent of space. Measurements obtained at similar locations will have similar error. Temporal autocorrelation implies that GPS measurement error is not independent of time. Measurements obtained at similar times will have a similar error due to similar atmospheric conditions and a similar satellite constellation \citep{bos_etal_2007}.  We carried out a simple experiment to visualize temporal autocorrelation in real-world GPS data. We placed a GPS unit at a known position $\boldsymbol{P}$ and recorded about 720 position estimates over a period of about six hours at a sampling rate of $1/30~\mathrm{Hz}$. The resulting distribution is centred around $\boldsymbol{P}$  with an $\text{\textit{R95}}$ of about $3~\mathrm{m}$~(Figure~\ref{fig:collage} a). If only those position estimates are displayed that were recorded within a certain time interval, GPS measurement error reveals itself to be highly auto-correlated. Figure~\ref{fig:collage} b, for example, shows only those position estimates that were obtained within periods covering $5$~minutes before and after ${t_1,t_2,t_3}$.

In this Section we build on the relationship described in Equation~\ref{eq:C} and show the spatial  and temporal autocorrelation in two sets of real-world GPS movement data. In the first experiment we identified to what degree a set of pedestrian movement data was affected by spatial and temporal autocorrelation.  In the second experiment we  derived the spatial autocorrelation in a set of car movement data. Based on this we tried to assess how well the GPS captured the movement of the car.

\subsection{Experiment 1: Pedestrian trajectories}

\subsubsection*{Experimental setup}

For the first experiment, we equipped a pedestrian with a GPS. The pedestrian walked along a reference course with a well-established reference distances $d_0$. The movement of the pedestrian was recorded with a QSTARZ:BT-Q1000X GPS unit\footnote{for specifications, please refer to: http://www.qstarz.com/Products/GPS/20Products/BT-Q1000.html} with `Assisted GPS' activated.

Rather than using a high-quality GPS we collected all data with a low-budget GPS, a type of GPS common for recording movement data. We deliberately treated the GPS as  a `black box'. This implies that the algorithm to calculate the position estimates from the raw GPS signal was not known. Moreover, we considered that it was sufficient to use only a single GPS unit, as the aim of the experiment was not to investigate the quality of the particular GPS, but to show the usefulness of our approach.

The reference course was located in an empty parking lot to avoid shadowing and multi-path effects. We staked out a square with sides that were $10~\mathrm{m}$ long and had markers at one meter intervals. A square was used in order to allow distance measurements to be collected in all four cardinal directions (approximately). The distance between the markers was used as a reference distance $d_0$.

The GPS position estimates were obtained by walking to the reference markers in turn and recording the position, moving around the square until all positions of the markers had been recorded. Position estimates were only taken at the reference markers, and only when the recording button was pushed manually. Two consecutive position estimates were taken within three to five seconds. A full circuit around the square took approximately between two and three minutes and resulted in $40$ positions being recorded. A total of $25$ circuits around the square were completed, without any breaks. This resulted in $1000$ GPS positions being collected in approximately one hour.

In pre-processing distance measurements $d^m$ were calculated between the position estimates and later compared to $d_0$ the reference distance between the markers. Then the average measured distance $\bar{d^m}$ was calculated and from this $\hat{\mathit{OD}} = \bar{d^m} -d_0$ and $\hat C =   d^2_0 - \bar{d^m_2}  +\textit{Var}_{\textit{gps}}$ were derived.  $\mathit{\hat {OD}}$  and $\hat C$ are estimators for $\mathit{OD}$ and $C$.

We decided not to derive $\sigma_X$ and $\sigma_Y$ from observational data, but to set $\sigma_X = \sigma_Y = 3\mathrm{m}$. Hence, $\textit{Var}_{\textit{gps}}$ is not the observed variance of GPS measurement error, but a reference value to which $\mathit{OD}$ is later compared to. Consequently, our results do not show the exact value of $C$, but provide an estimate of $C$ with respect to $\textit{Var}_{\textit{gps}}$.

We increased the spatial separation between two position estimates of the pedestrian to illustrate the influence of spatial autocorrelation. Then we increased the temporal separation between two position estimates to illustrate the influence of temporal autocorrelation.

\subsubsection*{Results}

\begin{figure}[]
	\centering
   \includegraphics[scale=0.6]{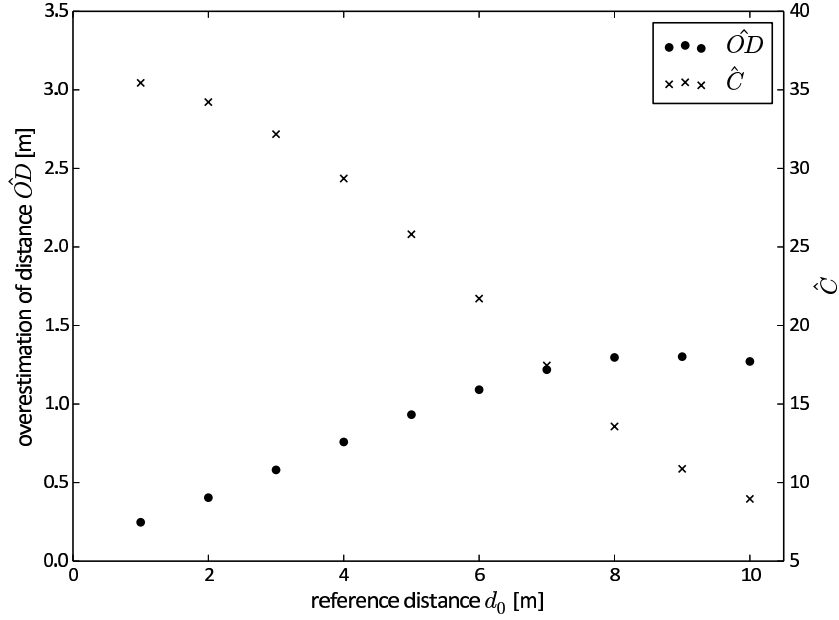} 
	\caption{Overestimation of distance ($\hat{\mathit{OD}}$) and spatial autocorrelation of GPS measurement error ($\hat C$) in the pedestrian movement data.}
	\label{fig:spatial_ac}
\end{figure}

In contrast to the theoretical findings in Figure~\ref{fig:OD_compared}, overestimation of distance tended to increase as the reference distance $d_0$ increased.  This was due to a decrease in the spatial autocorrelation of GPS measurement error. With increasing spatial separation of the position estimates,  measurement error became less auto-correlated. Figure~\ref{fig:spatial_ac} shows the relationship between the reference distance $d_0$  and $\hat{\mathit{OD}}$ (black dots) as well as $\hat{C}$ (black crosses). 

\begin{figure}[]
	\centering
   \includegraphics[scale=0.45]{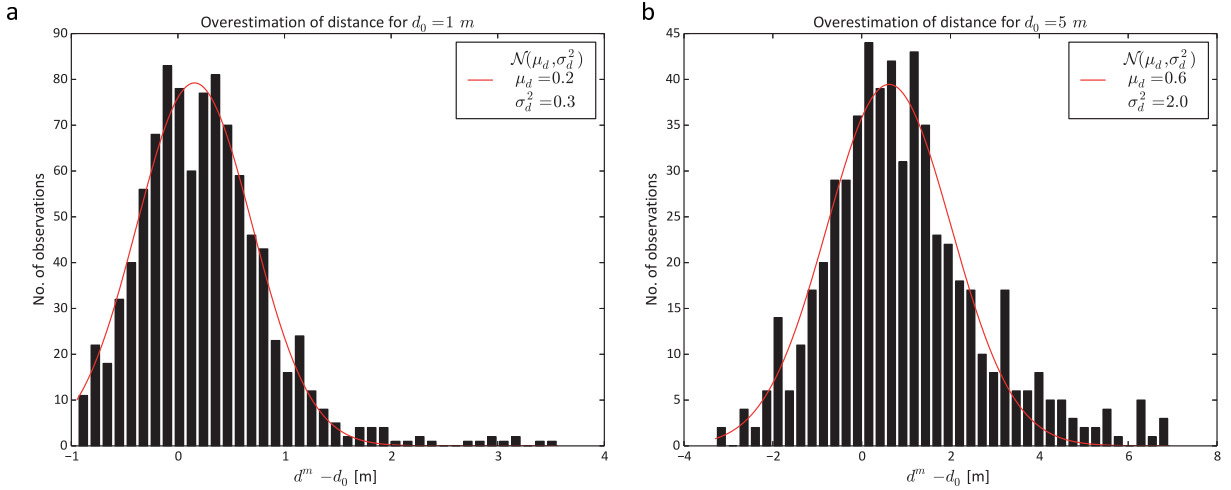} 
	\caption{Histogram of the difference between measured and reference distance ($d^m -d_0$) for  $d_0=1~\mathrm{m}$ (a) and $d_0=5~\mathrm{m}$ (b)}
	\label{fig:histogram}
\end{figure}

We wanted to illustrate that the overestimation of distance was not caused by a small number of extreme outliers. Figure~\ref{fig:histogram} shows the histogram of $d^m - d_0$  for $d_0=1~\mathrm{m}$ (a), and for $d_0=5~\mathrm{m}$ (b) and their  fit to a Gaussian distribution.  Both histograms follow a Gaussian distribution  $\mathcal{N} (\mu_{d} ,\sigma^2_{d}) $ rather well and outliers are almost non-existent. Note that $\mu_d$ and $\sigma^2_d$ in Figure~\ref{fig:histogram} refer to the values of the fitted Gaussian distribution and not to the empirically derived frequency.\\

\begin{figure}[]
	\centering
   \includegraphics[scale=0.5]{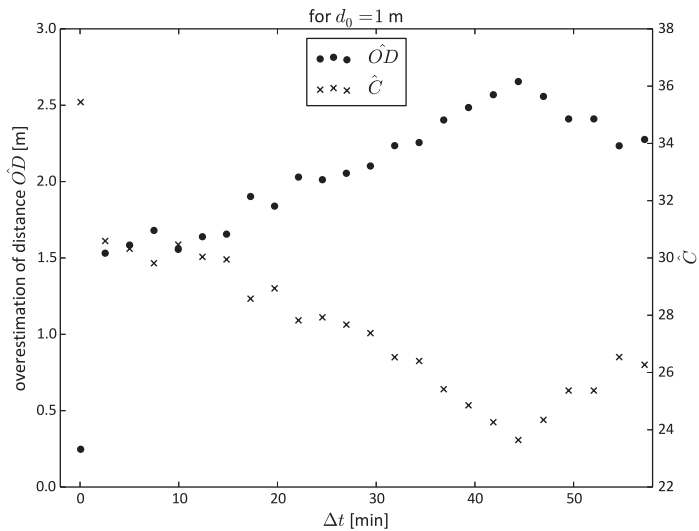} 
	\caption{Overestimation of distance $\hat{\mathit{OD}}$ and temporal autocorrelation of GPS measurement error ($\hat C$) in the pedestrian movement data.}
	\label{fig:temporal_ac}
\end{figure}

In order to illustrate the temporal autocorrelation in GPS measurement error, we calculated the distance between non-consecutive position estimates around the square. One example is the distance between two position estimates, where the second one was obtained one circuit after the first later. The reference distance between the markers remained the same, e.g. $ d_0 = 1~\mathrm{m}$, but the position estimates were recorded within a longer time interval $\Delta t$. Figure~\ref{fig:temporal_ac} shows the relationship between $\Delta t$ and $\hat{\mathit{OD}}$ (black dots) as well as $\hat{C}$ (black crosses) for a reference distance $ d_0= 1~\mathrm{m}$. $\hat{\mathit{OD}}$ increase with longer time intervals. The sharpest increase occurs between measurements that were taken promptly and those taken after about  $2 {\frac{1}{2}}$  minutes.  After $40$ minutes the curve levels out. This increase of $\hat{\mathit{OD}}$ was caused by the temporal auto-correlation of measurement error. For measurements taken within several seconds, measurement error appears to be strongly auto-correlated. However, auto-correlation falls sharply for measurements taken within $2 {\frac{1}{2}}$  minutes. From then on $\hat C$ gradually decreases as $\Delta t$ increases; again the curve levels out at about $40$ minutes.

The data for the above experiment were calculated with a GPS for which the algorithm to calculate the position estimates from the raw GPS signal was not known. This raises the legitimate question, whether the results were produced by a smoothing algorithm rather than the behaviour of the GPS. Let us assume that the GPS used a smoothing algorithm. In simplified form, the current position estimate is then calculated from the last position estimate, the current GPS measurement and a movement model. For movement with constant speed and direction, smoothing yields trajectories that represent the true movement very accurately. However, sudden changes in movement, i.e. a sharp turn, are not followed by the trajectory. The current measurement implies a sharp turn, however, the movement model  does not. Thus, the sharp turn becomes more elongated, the overestimation of distance increases. However, we did not find any support for an increase in the overestimation of distance after a sharp turn. This can also be seen in Figure~\ref{fig:collage} b.

\subsection{Experiment 2: Car trajectories}

In the first experiment the reference distance $d_0$ was staked out along a reference course. For obvious reasons this is not possible for recording the movement of a car. Hence we derived $d_0$ from speed measurements recorded with a car's controller area network bus (CAN bus). 

\subsubsection*{Experimental Setup}

We equipped a car with a GPS unit and tracked its movement for about 6 days. The car moved mostly in an urban road network at rather low speeds (average: $25~\mathrm{km/h}$).  The temporal sampling rate of recording was $1~\mathrm{Hz}$.  For the CAN bus measurements, a sensor recorded the rotation of the car's drive axle, from which $d_0$ was inferred. Thus $d_0$ is the distance travelled by the car according to the CAN bus. For the same phases of movement we compared $d_0$ to $d^m$, the distance travelled by the car according to the GPS position estimates.  As in the first experiment, we set $\sigma_X = \sigma_Y = 3\mathrm{m}$ and calculated  $\textit{Var}_{\textit{gps}}$.

The data were first pre-processed and cleaned. Parts with very high speed (above $ 140~\mathrm{km/h}$) and very rapid acceleration (above $5~\mathrm{m/s^2
}$)  were removed. Although the data consisted mostly of the car's forward movements, there were also periods when it was either stationary or reversing in a parking lot. The data may also have included some periods during which shadowing  caused a loss of the GPS signal (for example when driving in a tunnel). We therefore applied a simple mode detection algorithm to remove any such periods. The algorithm evaluates speed and acceleration along the trajectory and distinguishes segments that most probably reflect driving behaviour from those that are likely to reflect non-driving behaviour \citep{zheng_etal_2010}. Using the algorithm we were able to include only long phases of continuous driving, sampled at a continuous sampling frequency of $1~\rm{Hz}$. Following this pre-processing a total of about $ 195~\mathrm{km}$ of car trajectories remained for analysis.

\subsubsection*{Results}

\begin{figure}[]
	\centering
   \includegraphics[scale=0.5]{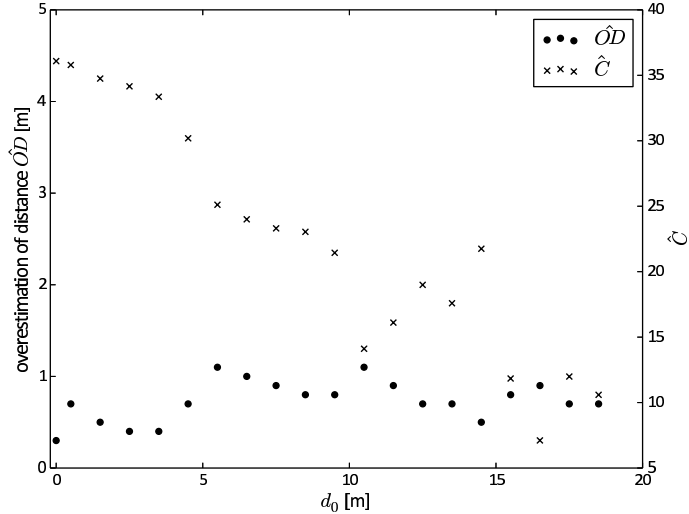} 
	\caption{Overestimation of distance ($\hat{\mathit{OD}}$) and spatial autocorrelation of GPS measurement error ($\hat C$) in the car movement data.}
	\label{fig:can_bus_autocorr}
\end{figure}

Figure~\ref{fig:can_bus_autocorr} shows that the autocorrelation of GPS measurement error decreased as the spatial separation between two consecutive position estimates increased. Nevertheless, $\hat C$ in Figure~\ref{fig:can_bus_autocorr} is always positive. This can be interpreted as a quality measure for the movement data. Consecutive position estimates were affected by less variance than initially suggested by $\textit{Var}_{\textit{gps}}$. 

Although the results in Figure~\ref{fig:can_bus_autocorr} are similar to those obtained from the pedestrian movement data, they contain outliers. We believe that these outliers occur due to two reasons.  First, the data comprise relatively few distance measurements for big $d_0$ because of the generally low speed of the car. Second, we could not guarantee a full temporal synchronization of both measurement systems (GPS and  CAN bus). In other words, $d_0$ and $d^m$ might relate to slightly different time intervals. We found this lag to be around one second. We believe that this insight is important for the practical application of Equation~\ref{eq:C}. In order to provide valid results it requires both a significant number of distance measurements as well as a proper synchronisation of reference and measured distance.

\section{Discussion and Outlook} \label{sec:discussion}

In this paper we identified a systematic bias in GPS movement data. If interpolation error can be neglected GPS trajectories systematically overestimate distances travelled by a moving object.  This overestimation of distance has previously been noted in the trajectories of fishing vessels \citep{palmer_2008}. For high sampling rates the distance travelled by the vessel was overestimated due to measurement error, while for lower sampling rates it was underestimated due to the influence of interpolation error. We provided a mathematical explanation for this phenomenon and showed that it functionally depends on three parameters, of which one is $C$, the spatio-temporal autocorrelation of GPS measurement error. We built on this relationship and introduced a novel approach to estimate $C$ in real-world GPS movement data. 
In this Section we want to discuss our findings and show their implications for movement analysis and beyond.  

In the era of big data, more and more movement data are recorded at finer and finer intervals. For movement recorded at very high frequencies (e.g. $1 \rm {Hz}$) interpolation error can usually be neglected. Hence $\mathit{OD}$ is bound to occur in these data.  However, this does not mean that  high frequency movement data are of low quality, quite the opposite is true. Using the relationship between $C$ and $\mathit{OD}$ we showed experimentally that GPS measurement error in real world trajectories was affected by strong spatial and temporal autocorrelation. In other words, if the data were recorded close in space and time they captured the movement of the object better than if they were further apart.

Autocorrelation is important for movement analysis in many aspects. An appropriate sampling strategy for recording movement data, for example, should consider the influence of measurement error and address spatial and temporal autocorrelation. Since autocorrelation can be interpreted as a quality measure, it allows to reveal the performance  of different GPS receivers in different recording environments.  Moreover, autocorrelation has implications for simulation. \cite{laube_etal_2011} performed a simulation to reveal the complex interaction between measurement error and interpolation error and their effects on recording speed, turning angle and sinuosity. Their  Monte Carlo simulation assumed GPS errors to scatter entirely randomly between each two consecutive positions. Our approach allows to verify whether this assumption is realistic.

\subsubsection* {Where to find a reference distance?}

For practical applications the biggest limitation of our experiments is their dependency on a valid reference distance. The moving object must traverse the reference distance along a straight line and without interpolation error, and at a precisely known time. Moreover, a large number of measurements has to be collected, since $C$ is derived from the expectation value of a random variable.  

This limitation leads to a possibly interesting application of our findings, where the reference distance is derived from the GPS point speed measurements. Point speed measurements are calculated from the instantaneous derivative of the GPS signal using the Doppler effect. Point speed is very accurate \citep{bruton_etal_1999} and usually part of a GPS position estimate. Hence, for high sampling rates (e.g.$1$ \rm {Hz}) point speed measurements can be used to infer the distance that a moving object has travelled between two position estimates. This distance is not affected by the overestimation of distance effect and could serve as a reference distance. Thus, GPS could be compared to itself to reveal the spatio-temporal autocorrelation of the position estimates. This approach would not require any other ground truth data. However, its feasibility and usefulness are yet to be tested.  

Finally, we want to underline that our findings are not only relevant for GPS. The overestimation of distance is bound to occur in any type of movement data where distances are deduced from imprecise position estimates, of course only if interpolation error can be neglected.

\section*{Acknowledgments}

This research was funded by the Austrian Science Fund (FWF) through the Doctoral College GIScience at the University of Salzburg (DK W 1237-N23). We thank Arne Bathke from the Department of Mathematics of the University of Salzburg for his invaluable help on quadratic forms.

\bibliography{tGISguide.bib}

\newcommand{\noopsort}[1]{} \newcommand{\printfirst}[2]{#1}
  \newcommand{\singleletter}[1]{#1} \newcommand{\switchargs}[2]{#2#1}
\begin{thebibliography}{35}
\providecommand{\natexlab}[1]{#1}

\bibitem[\protect\citeauthoryear{Andrienko
  {\itshape{et~al.}}}{2011}]{andrienko_etal_2011a}
Andrienko, G., Andrienko, N., and Heurich, M., 2011. An event-based conceptual
  model for context-aware movement analysis. {\itshape International Journal of
  Geographical Information Science}, 25 (9), 1347--1370.

\bibitem[\protect\citeauthoryear{Bos {\itshape{et~al.}}}{2008}]{bos_etal_2007}
Bos, M., {\itshape et~al.}, 2008. Fast error analysis of continuous {GPS}
  observations. {\itshape Journal of Geodesy}, 82 (3), 157--166.

\bibitem[\protect\citeauthoryear{Bruton
  {\itshape{et~al.}}}{1999}]{bruton_etal_1999}
Bruton, A., Glennie, C., and Schwarz, K., 1999. Differentiation for
  high-precision {GPS} velocity and acceleration determination. {\itshape GPS
  solutions}, 2 (4), 7--21.

\bibitem[\protect\citeauthoryear{Chin}{1987}]{chin_1987}
Chin, G.Y., 1987. {\itshape {Two-dimensional Measures of Accuracy in
  Navigational Systems}. }Report DOT-TSC-RSPA-87-1, U.S. Department of
  Transportation.

\bibitem[\protect\citeauthoryear{Codling
  {\itshape{et~al.}}}{2008}]{codling_etal_2008}
Codling, E.A., Plank, M.J., and Benhamou, S., 2008. Random walk models in
  biology. {\itshape Journal of the Royal Society Interface}, 5 (25), 813--834.

\bibitem[\protect\citeauthoryear{Colombo}{1986}]{colombo_1986}
Colombo, O.L., 1986. Ephemeris errors of {GPS} satellites. {\itshape Bulletin
  g{\'e}od{\'e}sique}, 60 (1), 64--84.

\bibitem[\protect\citeauthoryear{D'Eon
  {\itshape{et~al.}}}{2002}]{deon_etal_2002}
D'Eon, R.G., {\itshape et~al.}, 2002. {GPS} radiotelemetry error and bias in
  mountainous terrain. {\itshape Wildlife Society Bulletin},  430--439.

\bibitem[\protect\citeauthoryear{Douglas-Hamilton
  {\itshape{et~al.}}}{2005}]{douglas-hamilton_etal_2005}
Douglas-Hamilton, I., Krink, T., and Vollrath, F., 2005. Movements and
  corridors of African elephants in relation to protected areas. {\itshape
  Naturwissenschaften}, 92 (4), 158--163.

\bibitem[\protect\citeauthoryear{El-Rabbany and
  Kleusberg}{2003}]{el-rabbany_kleusberg_2003}
El-Rabbany, A. and Kleusberg, A., 2003. Effect of temporal physical correlation
  on accuracy estimation in {GPS} relative positioning. {\itshape Journal of
  Surveying Engineering}, 129 (1), 28--32.

\bibitem[\protect\citeauthoryear{{G\"u}ting and
  Schneider}{2005}]{gueting_schneider_2005}
{G\"u}ting, R. and Schneider, M., 2005. {\itshape Moving objects databases}.
  San Francisco: Morgan Kaufmann.

\bibitem[\protect\citeauthoryear{Higuchi and
  Pierre}{2005}]{higuchi_pierre_2005}
Higuchi, H. and Pierre, J.P., 2005. Satellite tracking and avian conservation
  in Asia. {\itshape Landscape and Ecological Engineering}, 1 (1), 33--42.

\bibitem[\protect\citeauthoryear{Hofmann-Wellenhof
  {\itshape{et~al.}}}{2003}]{hofmann-wellenhof_etal_2003}
Hofmann-Wellenhof, B., Legat, K., and Wieser, M., 2003. {\itshape Navigation:
  {Principles} of positioning and guidance}.   Wien: Springer Verlag.

\bibitem[\protect\citeauthoryear{Howind
  {\itshape{et~al.}}}{1999}]{howind_etal_1999}
Howind, J., Kutterer, H., and Heck, B., 1999. Impact of temporal correlations
  on {GPS}-derived relative point positions. {\itshape Journal of Geodesy}, 73
  (5), 246--258.

\bibitem[\protect\citeauthoryear{Jerde and
  Visscher}{2005}]{jerde_visscher_2005}
Jerde, C.L. and Visscher, D.R., 2005. {GPS} measurement error influences on
  movement model parameterization. {\itshape Ecological Applications}, 15 (3),
  806--810.

\bibitem[\protect\citeauthoryear{Jun {\itshape{et~al.}}}{2006}]{jun_etal_2006}
Jun, J., Guensler, R., and Ogle, J.H., 2006. Smoothing methods to minimize
  impact of {Global Positioning System} random error on travel distance, speed,
  and acceleration profile estimates. {\itshape Transportation Research Record:
  Journal of the Transportation Research Board}, 1972 (1), 141--150.

\bibitem[\protect\citeauthoryear{Klenke}{2013}]{klenke_2013}
Klenke, A., 2013. {\itshape Probability theory: a comprehensive course}.
  Springer Science \& Business Media.

\bibitem[\protect\citeauthoryear{Langley}{1997}]{langley_1997}
Langley, R.B., 1997. The {GPS} error budget. {\itshape GPS world}, 8 (3),
  51--56.

\bibitem[\protect\citeauthoryear{Laube and Purves}{2011}]{laube_etal_2011}
Laube, P. and Purves, R.S., 2011. How fast is a cow? {Cross-Scale Analysis of
  Movement Data}. {\itshape Transactions in GIS}, 15 (3), 401--418.

\bibitem[\protect\citeauthoryear{Macedo
  {\itshape{et~al.}}}{2008}]{macedo_etal_2008}
Macedo, J., {\itshape et~al.}, 2008. 5. {\itshape {In}}: {\itshape Trajectory
  Data Models}.,  123 -- 150  Berlin: Springer.

\bibitem[\protect\citeauthoryear{Mintsis
  {\itshape{et~al.}}}{2004}]{mintsis_etal_2004}
Mintsis, G., {\itshape et~al.}, 2004. Applications of GPS technology in the
  land transportation system. {\itshape European Journal of Operational
  Research}, 152 (2), 399--409.

\bibitem[\protect\citeauthoryear{Modsching
  {\itshape{et~al.}}}{2006}]{modschnig_etal_2006}
Modsching, M., Kramer, R., and ten Hagen, K., 2006. Field trial on {GPS
  Accuracy} in a medium size city: {The} influence of built-up. {\itshape
  {In}}:  {\itshape 3rd Workshop on Positioning, Navigation and Communication},
   209--218.

\bibitem[\protect\citeauthoryear{Olynik}{2002}]{olynik_2002}
Olynik, M., 2002. {\itshape Temporal characteristics of {GPS} error sources and
  their impact on relative positioning. }Report, University of Calgary.

\bibitem[\protect\citeauthoryear{Palmer}{2008}]{palmer_2008}
Palmer, M.C., 2008. Calculation of distance traveled by fishing vessels using
  {GPS} positional data: {A} theoretical evaluation of the sources of error.
  {\itshape Fisheries Research}, 89 (1), 57--64.

\bibitem[\protect\citeauthoryear{Parent
  {\itshape{et~al.}}}{2013}]{parent_etal_2013}
Parent, C., {\itshape et~al.}, 2013. Semantic trajectories modeling and
  analysis. {\itshape ACM Computing Surveys}, 45 (4), 1--32.

\bibitem[\protect\citeauthoryear{Schneider}{1999}]{schneider_1999}
Schneider, M., 1999. {\itshape {In}}: {\itshape Uncertainty management for
  spatial data in databases: {Fuzzy} spatial data types}.,  330--351  Berlin:
  Springer.

\bibitem[\protect\citeauthoryear{Sigrist
  {\itshape{et~al.}}}{1999}]{sigrist_etal_1999}
Sigrist, P., Coppin, P., and Hermy, M., 1999. Impact of forest canopy on
  quality and accuracy of {GPS} measurements. {\itshape International Journal
  of Remote Sensing}, 20 (18), 3595--3610.

\bibitem[\protect\citeauthoryear{Van~der Spek
  {\itshape{et~al.}}}{2009}]{vanderspek_etal_2009}
Van~der Spek, S., {\itshape et~al.}, 2009. Sensing human activity: GPS
  tracking. {\itshape Sensors}, 9 (4), 3033--3055.

\bibitem[\protect\citeauthoryear{Wang
  {\itshape{et~al.}}}{2002}]{wang_etal_2002}
Wang, J., Satirapod, C., and Rizos, C., 2002. Stochastic assessment of {GPS}
  carrier phase measurements for precise static relative positioning. {\itshape
  Journal of Geodesy}, 76 (2), 95--104.

\bibitem[\protect\citeauthoryear{Wang
  {\itshape{et~al.}}}{2014}]{wang_etal_2014}
Wang, J., {\itshape et~al.}, 2014. A novel approach for generating routable
  road maps from vehicle GPS traces. {\itshape International Journal of
  Geographical Information Science}, 29 (1), 69--91.

\bibitem[\protect\citeauthoryear{{William J. Hughes Technical
  Center}}{2013}]{gps_report_2013}
{William J. Hughes Technical Center}, 2013. {\itshape {Global Positioning
  System (GPS) Standard Positioning Service (SPS) Performance Analysis Report
  }. }Report, Federal Aviation Administration.

\bibitem[\protect\citeauthoryear{Wing
  {\itshape{et~al.}}}{2005}]{wing_etal_2005}
Wing, M.G., Eklund, A., and Kellogg, L.D., 2005. Consumer-grade global
  positioning system {(GPS)} accuracy and reliability. {\itshape Journal of
  Forestry}, 103 (4), 169--173.

\bibitem[\protect\citeauthoryear{Zandbergen}{2009}]{zandbergen_2009}
Zandbergen, P.A., 2009. Accuracy of {iPhone} locations: {A comparison of
  assisted GPS, WiFi} and cellular positioning. {\itshape Transactions in GIS},
  13 (s1), 5--25.

\bibitem[\protect\citeauthoryear{Zheng
  {\itshape{et~al.}}}{2010}]{zheng_etal_2010}
Zheng, Y., {\itshape et~al.}, 2010. Understanding transportation modes based on
  {GPS} data for web applications. {\itshape ACM Transactions on the Web
  (TWEB)}, 4 (1), 1.

\bibitem[\protect\citeauthoryear{Zheng
  {\itshape{et~al.}}}{2011}]{zheng_etal_2011a}
Zheng, Y., {\itshape et~al.}, 2011. Urban computing with taxicabs. {\itshape
  {In}}:  {\itshape Proceedings of the 13th international conference on
  Ubiquitous computing},  89--98.

\bibitem[\protect\citeauthoryear{Zito
  {\itshape{et~al.}}}{1995}]{zito_etal_1995}
Zito, R., d'Este, G., and Taylor, M.A., 1995. Global positioning systems in the
  time domain: How useful a tool for intelligent vehicle-highway systems?.
  {\itshape Transportation Research Part C: Emerging Technologies}, 3 (4),
  193--209.

\end{thebibliography}

\end{document}